\documentclass[aps,pre,twocolumn,english,superscriptaddress,showpacs,floatfix]{revtex4}
\usepackage{amsmath}
\usepackage{psfig,graphics,graphicx}
\usepackage{babel}

\begin{document}

\title{Front explosion in a periodically forced surface reaction}
\author{J\"orn Davidsen}
\email{[]davidsen@pks.mpg.de}
\affiliation{Max-Planck-Institut f\"ur Physik Komplexer Systeme, N\"othnitzer Strasse 38,
01187 Dresden, Germany}
\author{Alexander Mikhailov}
\affiliation{Fritz-Haber-Institut der Max-Planck-Gesellschaft, Faradayweg 4-6, 14195
Berlin, Germany}
\author{Raymond Kapral}
\affiliation{Fritz-Haber-Institut der Max-Planck-Gesellschaft, Faradayweg 4-6, 14195
Berlin, Germany}
\affiliation{Chemical Physics Theory Group, Department of Chemistry, University of
Toronto, Toronto, ON M5S 3H6, Canada}
\date{\today}

\begin{abstract}
Resonantly-forced oscillatory reaction-diffusion systems can exhibit fronts
with complicated interfacial structure separating phase-locked homogeneous
states. For values of the forcing amplitude below a critical value the front
``explodes" and the width of the interfacial zone grows without bound. Such
front explosion phenomena are investigated for a realistic model of
catalytic CO oxidation on a Pt(110) surface in the 2:1 and 3:1
resonantly-forced regimes. In the 2:1 regime, the fronts are stationary and
the front explosion leads to a defect-mediated turbulent state. In the 3:1
resonantly-forced system, the fronts propagate. The front velocity tends to
zero as the front explosion point is reached and the final asymptotic state
is a 2:1 resonantly-locked labyrinthine pattern. The front dynamics
described here should be observable in experiment since the model has been
shown to capture essential features of the CO oxidation reaction.
\end{abstract}


\maketitle

\section{Introduction}

Catalytic surface reactions are a large class of chemical reactions with
important applications. They are often accompanied by oscillations, and a
rich variety of wave patterns, as well as chemical turbulence, can be
observed in such systems \cite{Ertl,Rotermund}. An extensively studied
surface reaction is CO oxidation on single crystals of platinum. When the
reaction takes place on the crystallographic plane Pt(110), a structural
phase transition in the top layer of the substrate is coupled to the
reaction, making oscillations and excitability possible. These kinetic
regimes are well reproduced by the theoretical model of Krischer, Eiswirth
and Ertl \cite{Krischer} where reaction-induced surface reconstruction is
taken into account. The bifurcation analysis of this model shows that it has
a supercritical Hopf bifurcation \cite{Krischer}. Moreover, in a region of
parameter space the uniform oscillations in this system may be unstable with
respect to the modulational Benjamin-Feir instability, leading to the
spontaneous development of turbulence \cite{Falcke}. Therefore, in the
vicinity of the Hopf bifurcation, the CO oxidation reaction approximately
obeys the complex Ginzburg-Landau equation (CGLE) \cite{Kuramoto}.
Experimentally, it may be difficult to maintain the reaction so close to the
bifurcation line that the oscillations have a harmonic character and their
amplitude is small. Nonetheless, many qualitative features of the wave
patterns typically found in the CGLE are also observed in the CO oxidation
reaction on Pt(110) under oscillatory conditions. In particular, the
spontaneous development of amplitude turbulence through the nucleation of
spiral wave pairs has been found in this system \cite%
{Science,BertramPREexp,BertramJPC}. Thus, the CO oxidation reaction on
Pt(110) provides an opportunity to experimentally investigate the general
phenomena of spatiotemporal chaos in weakly nonlinear oscillatory
reaction-diffusion systems.

To control the CO oxidation reaction, the partial pressure of CO in the gas
above the catalytic surface can be modulated in time by varying the dosing
rate of CO into the reaction chamber. The modulation of the CO pressure
changes reaction conditions in the same way in all surface elements and,
therefore, its effect is global. By making the modulation dependent on the
properties of monitored patterns, global feedback can be introduced in the
reaction system. Suppression of chemical turbulence and development of new,
feedback-induced wave patterns, including regimes of intermittent
spatiotemporal chaos, have been experimentally observed \cite%
{Science,BertramPREexp,Beta} and theoretically investigated \cite%
{Science,BertramPRETheor} for the CO oxidation reaction.

Alternatively, external periodic modulation of the partial CO
pressure can be applied, leading to a periodic forcing of this
chemical reaction. In experiments \cite{Eiswirth1988}, entrainment
of CO oscillations by such periodic forcing has been detected, but
spatially-resolved observations of concentration wave patterns on
the platinum surface were not then possible. Subsequently,
spatially-resolved experiments on periodic forcing of chemical
turbulence in this reaction were performed \cite{BertramJPC}. In
addition to full entrainment with uniform oscillations, regimes
with oscillating cells or labyrinthine patterns, and cascades of
amplitude defects characteristic of spatiotemporal intermittency,
were observed. Effects of periodic forcing have also been
experimentally investigated in the oscillatory
Belousov-Zhabotinsky reaction \cite{petrovnature,linPRE,linPRL},
where uniform oscillations are, however, always stable and
spontaneous development of turbulence is not known.

Previous theoretical studies \cite{faradaypaper,physicaDpaper} of fronts
separating different phase-locked states in the periodically forced CGLE
have shown that, under the conditions of the Benjamin-Feir instability, such
fronts may undergo "explosions" as the forcing amplitude is decreased. When
such an explosion occurs, the interfacial region separating the two locked
states grows and eventually fills the entire medium with a turbulent phase.
Similar phenomena were also found in coupled map lattice models \cite%
{klop,klp}.

In this paper, the effects of front explosions under 2:1 and 3:1 resonant
forcing are studied in the realistic model of CO oxidation on Pt(110) for
parameter values typical for the experiments on this chemical reaction where
non-harmonic oscillations with substantial amplitudes are observed. As we
show, front explosions occur in very different ways in these two resonant
regimes. Our work provides a framework for future experimental studies of
front explosions in the CO oxidation reaction.

\section{Periodically Forced Surface Reaction}

\label{sec:model}

\subsection*{Model}

We consider the realistic model \cite{Krischer} of catalytic CO oxidation on
a Pt(110) surface. The model takes into account adsorption of CO and O$_{2}$
molecules, reaction, desorption of CO molecules, the structural phase
transition of the Pt(110) surface, and the surface diffusion of adsorbed CO
molecules. Letting $u(\mathbf{r})$ and $v(\mathbf{r})$ represent the surface
coverage of CO and O at position $\mathbf{r}$, respectively, and $w(\mathbf{r%
})$ the local fraction of the surface area found in the non-reconstructed $%
1\times 1$ form, the reaction-diffusion equations giving the time evolution
of these fields are
\begin{eqnarray}
u_{t} &=&k_{1}\,s_{CO}\,p_{CO}(1-u^{3})-k_{2}\,u-k_{3}\,u\,v+D\nabla ^{2}u
\notag \\
v_{t} &=&k_{4}\,p_{O_{2}}[s_{O,1\times 1}\,w+s_{O,1\times 2}(1-w)](1-u-v)^{2}
\notag \\
&&-k_{3}\,u\,v)  \notag \\
w_{t} &=&k_{5}\left( \frac{1}{1+\exp (\frac{u_{0}-u}{\delta u})}-w\right)\;.
\end{eqnarray}%
All three fields can vary in the interval from $[0,1]$ and we define $%
\mathbf{c}(\mathbf{r},t)=(u(\mathbf{r},t),v(\mathbf{r},t),w(\mathbf{r},t))$
for future reference. For an explanation and specification of the values of
the parameters see Table~\ref{tab1}. Although certain features, such as
surface roughening, faceting, formation of subsurface oxygen, and the
effects of internal gas-phase coupling are not taken into account, this
model has proven to be remarkably successful in describing most aspects of
the experimental observations on this system. In particular, it is able to
capture oscillations in the CO oxidation process seen under some
experimental conditions.
\begin{table}[tbp]
\caption{Parameters of the model.}
\label{tab1}%
\begin{ruledtabular}
\begin{tabular}{lcc}
$k_1$ & $3.14\times10^5 \, \mbox{s}^{-1} \, \mbox{mbar}^{-1}$ & Impingement rate of CO\\
$k_2$ & $10.21 \, \mbox{s}^{-1}$ & CO desorption rate\\
$k_3$ & $283.8 \, \mbox{s}^{-1}$ & Reaction rate\\
$k_4$ & $5.860\times10^5 \, \mbox{s}^{-1} \, \mbox{mbar}^{-1}$ & Impingement rate of O$_2$\\
$k_5$ & $1.610 \, \mbox{s}^{-1}$ & Phase transition rate \\
$s_{CO}$ & 1.0 & CO sticking coefficient\\
$s_{O,1\times1}$ & 0.6 & Oxygen sticking coefficient\\
& & on the $1 \times 1$ phase\\
$s_{O,1\times2}$ & 0.4 & Oxygen sticking coefficient\\
& & on the $1 \times 2$ phase\\
$u_0, \delta u$ & 0.35, 0.05 & Parameters for the\\
& & structural phase transition\\
$D$ & $40 \, \mu \mbox{m}^2 \, \mbox{s}^{-1}$ & CO diffusion coefficient\\
$p_{O_2}$ & $12.0\times10^{-5} \, \mbox{mbar}$ & O$_2$ partial pressure\\
$p_0$ & $4.6219548\times10^{-5} \, \mbox{mbar}$ & Base CO partial pressure
\end{tabular}
\end{ruledtabular}
\end{table}

We are interested in situations where the catalytic surface reaction is
subjected to external periodic forcing. Experimentally, it is convenient to
periodically modulate the CO partial pressure $p_{CO}$. Consequently, we
assume in this paper that $p_{CO}$ varies according to the equation
\begin{equation}
p_{CO}(t) = p_0 (1 - {\ a \cos{\omega_f \, t}}),
\end{equation}
where $a$ is the amplitude of the forcing and $\omega_f$ the frequency of
the forcing. The base value of partial CO pressure is $p_0$. In the present
study the parameter values of the partial pressures are chosen such that in
the absence of forcing uniform oscillations are unstable with respect to
small perturbations and chemical turbulence spontaneously develops in the
unperturbed system. The reaction itself is oscillatory with period $T_0 =
2.550049\, $s for $a=0$.

Numerical simulations of the model were performed using a first-order finite
difference scheme for the spatial discretization with a grid resolution of $%
\Delta x = 4 \mu$m. For the temporal discretization an explicit Euler scheme
with a fixed time step $\Delta t = 0.0001$s was used. The system size was $%
L^2 = 0.4 \times 0.4$mm$^2$ and no-flux boundary conditions were imposed.

\subsection*{Fronts separating resonantly locked states}

We suppose that when periodic forcing is applied to the system it is locked
in an $n:m$ resonance. In this case there are $n$ distinct phases of the
oscillation. Chemical fronts separating pairs of these $n$ resonantly locked
states can be formed by a suitable choice of initial conditions. In
particular, in a two-dimensional domain, we choose initial conditions such
that the left and the right half planes are homogeneous but in two different
phases, respectively, and separated by a small interfacial zone where the
field values are chosen at random.

\begin{figure}[htbp]
\includegraphics*[width=\columnwidth]{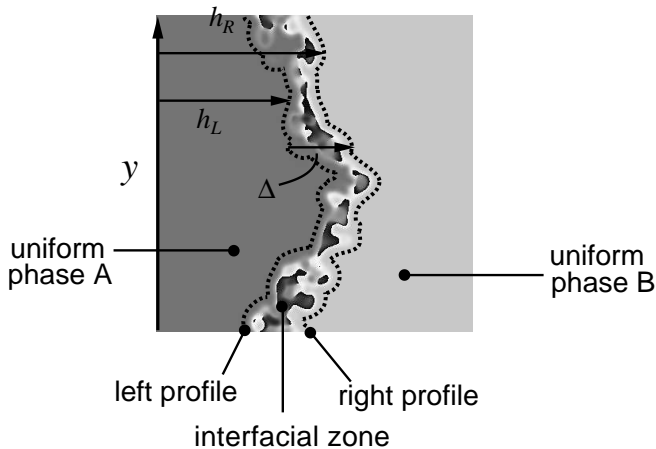}
\caption{Sketch of a phase front separating homogeneous domains of two
resonantly locked phases. The interfacial zone is delimited by left and
right profiles, $h_L$ and $h_R$, respectively. }
\label{phase_front}
\end{figure}

Under conditions where the unforced system is Benjamin-Feir unstable and the
forcing amplitude is sufficiently large, such initial conditions generate a
front separating the homogeneous phases which has the form of an interfacial
zone delineated by left and right profiles. The situation is depicted in
Fig.~\ref{phase_front}.
We let $h_L(y,t) = \mathrm{min}\{x : |\mathbf{c}(x,y,t) - \mathbf{c}^L_0(t)|
\geq \epsilon\}$ and $h_R(y,t) = \mathrm{max}\{x : |\mathbf{c}(x,y,t) -
\mathbf{c}^R_0(t)| \geq \epsilon\}$ (see Fig.~\ref{phase_front}). Here $%
\mathbf{c}^R_0(t)$ and $\mathbf{c}^L_0(t)$ are the concentration fields in
the two mode locked states that the front separates and $\epsilon$ is a
small numerical value which is chosen to be 0.01 in the simulations
presented below. We can thus define the interfacial zone as $I(t) = \{(x,y):
h_L(y,t) \le x \le h_R(y,t)\}$.

The dynamics of the front can be analyzed in terms of several quantities
that characterize its structure. The instantaneous intrinsic thickness of
the interfacial zone at position $y$ along the front is given by
\begin{equation}
\Delta(y,t) = h_R(y,t) - h_L(y,t).
\end{equation}
The spatial average $\langle\Delta\rangle(t)$ of the intrinsic thickness may
be computed from
\begin{equation}
\langle\Delta\rangle(t) = \frac{1}{L} \int^L_0 \mbox{d}y \, \Delta(y,t),
\end{equation}
where, henceforth, the angle brackets will refer to a spatial average along $%
y$. We also define the instantaneous position of the front at a point $y$
along the front by
\begin{equation}
x_f(y,t) = (h_R(y,t) + h_L(y,t))/2,
\end{equation}
and the instantaneous mean position as $\langle x_f \rangle(t)$. After a
transient time, the phase front dynamics is observed to enter a
statistically stationary regime where the temporal average of the time
derivatives of $\langle x_f \rangle(t)$ and $\langle\Delta\rangle(t)$ are
independent of time. These quantities are called the front velocity $v_f$
and the interface growth rate $v_\Delta$, respectively. If $v_\Delta \equiv
0 $ then the interface has an average finite thickness which is denoted by $%
\Delta_0$.

Studies of the 3:1 resonantly forced complex Ginzburg-Landau
equation have shown that if the amplitude of the forcing is
reduced, a critical value is reached where the average thickness
of the interfacial zone $\Delta_0$ is no longer finite. For such
values of the forcing, when the system starts from the two-phase
initial conditions described above, the interfacial zone grows to
fill the entire domain. We term this phenomenon a ``front
explosion". We now investigate this phenomenon for the resonantly
forced catalytic surface oxidation reaction.

\section{2:1 forcing}

\label{sec:2:1}

\begin{figure}[tbph]
\includegraphics*[width=\columnwidth,angle=180]{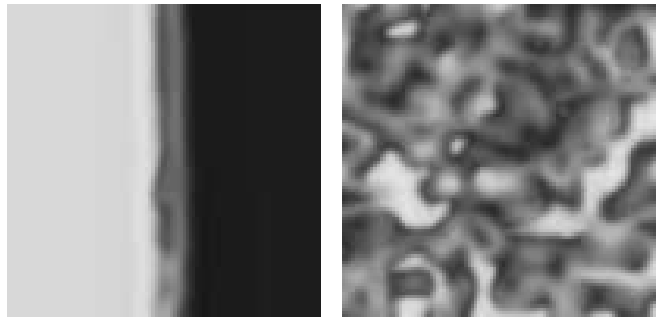}
\caption{Snapshots of $u(\mathbf{r})$ in the asymptotic state for $%
T_{f}=1.275$s without front explosion (left, $a=0.0138$) and with front
explosion (right, $a=0.00864$). Black corresponds to small values of $u$ and
white to large values of $u$. }
\label{asymp_state}
\end{figure}

\begin{figure}[tbph]
\includegraphics*[width=\columnwidth]{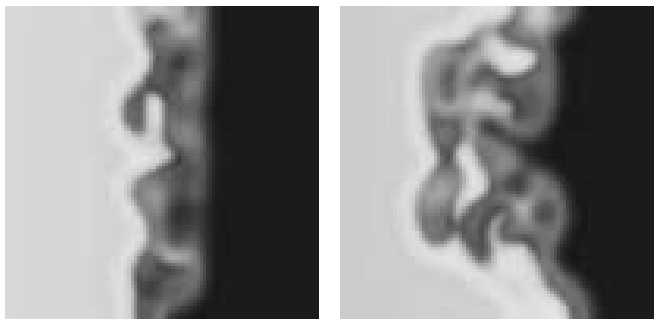}
\caption{Snapshot series of $u(\mathbf{r})$ for $a=0.0130$ showing a typical
front explosion. Left, $t=5\times T_{f}$; right, $t=20\times T_{f}$ with $%
T_{f}=1.275$s. Color coding as in Fig.~\protect\ref{asymp_state}. }
\label{explosion}
\end{figure}

In this section we study the front dynamics when the system is close to the
2:1 resonance, i.e., $T_{f}=T_{0}/2$ and, hence, $\omega _{f}=2\omega _{0}$.
In the 2:1 resonance regime we have two distinct equivalent phases that
differ by a phase shift of $\pi $. An example of a front separating the two
phase states is shown in the left panel of Fig.~\ref{asymp_state}.
The front is stationary with finite average width $\Delta _{0}$. As the
amplitude of the forcing is decreased we observe that $\Delta _{0}$ grows
and, for amplitudes lower than a critical value, the interfacial zone grows
without bound and the front explodes. An example of an exploding front is
shown in Fig.~\ref{explosion}.
In this case the turbulent zone grows to fill the finite domain leading to
turbulent dynamics in the entire system as shown in the right panel of Fig.~%
\ref{asymp_state}.

\begin{figure}[htbp]
\includegraphics*[width=\columnwidth]{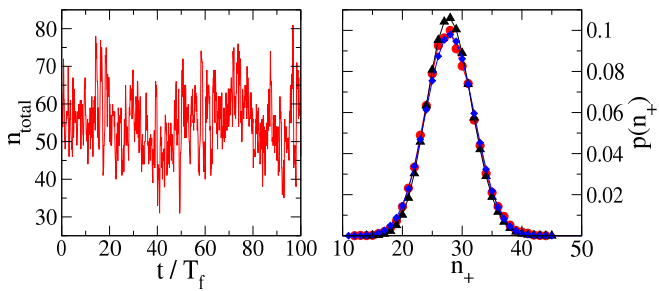}
\caption{(Color online) Left: Number of defects $n_{total}$ as a function of
time (same forcing as in the right panel of Fig.~\protect\ref{asymp_state})
for $\protect\phi (\mathbf{r},t) = \arctan [(w(\mathbf{r},t) - w^0) / (u(%
\mathbf{r},t) - u^0)]$ with $(w^0, u^0) = (0.354, 0.487)$ as the center of
rotation. Right: Probability distribution of the number of defects with
positive topological charge $n_+$ (red circles). It is clearly different
from a squared Poissonian (black triangles) \protect\cite{gil90} but shows
very good agreement with a Gaussian (blue diamonds). }
\label{defect_all}
\end{figure}

The turbulent state that emerges when the interfacial zone explodes has the
characteristics of defect-mediated turbulence, where the dynamics of a
pattern is dominated by the rapid motion, nucleation and annihilation of
point defects (vortices or dislocations) \cite{coullet89}. A defect is
characterized by its integer topological charge (or winding number) $m_{top}$
which is defined by $\frac{1}{2\pi }\oint \nabla \phi (\mathbf{r},t)\cdot d%
\mathbf{l}=\pm m_{top}$ \cite{mermin79}, where $\phi (\mathbf{r},t)$ is the
local phase and the integral is taken along a closed curve surrounding the
defect. A topological defect, thus, corresponds to a point in the medium
where the local amplitude is zero and the phase is not defined. In the CGLE
only topological defects with $m_{top} = \pm 1$ are observed and that is
also the case here. In Fig.~\ref{defect_all} (left panel) we see that the
number of defects fluctuates about a statistically stationary average value.
The probability distribution of defects with positive topological charge
shown in the right panel of this figure has an approximate Gaussian form.
This is in accord with a simple model of defect dynamics based on rate
equations and taking into account no-flux boundary conditions \cite%
{gil90,daniels02}. The model predicts a modified Poissonian distribution
which, in the limit of large mean defect numbers, converges to a Gaussian.

\begin{figure}[htbp]
\includegraphics*[width=\columnwidth]{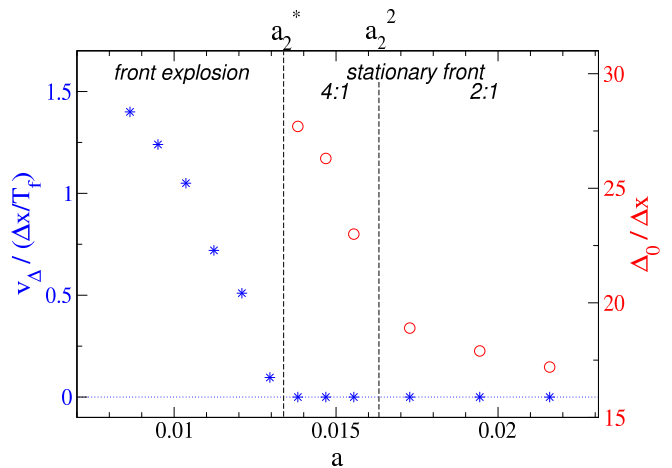}
\caption{(Color online) Plot of the interface width $\Delta_0$ (red circles)
and the interface growth rate $v_{\Delta}$ (blue stars) as a function of the
forcing amplitude $a$ for $T_f = 1.275$s. The statistical errors are less
than the size of the symbols. Three different regimes can be identified. See
text for details. }
\label{front_diagram2-1}
\end{figure}

A quantitative characterization of the transition from a stationary front
with finite width ($v_f = v_\Delta \equiv 0$) to an exploding front ($v_f
\equiv 0$ and $v_\Delta > 0$) is given in Fig.~\ref{front_diagram2-1} which
plots both $\Delta_0$ and $v_{\Delta}$ versus the forcing amplitude $a$.
The transition occurs at a critical value $a_2^*$ which coincides, within
the numerical resolution, with a bifurcation leading to the coexistence and
bistability of homogeneous oscillatory states and turbulent states. The
homogeneous oscillatory states become unstable only for values of $a$ much
smaller than $a_2^*$, namely at $a\approx0.0006$.

\begin{figure}[tbp]
\includegraphics*[width=\columnwidth]{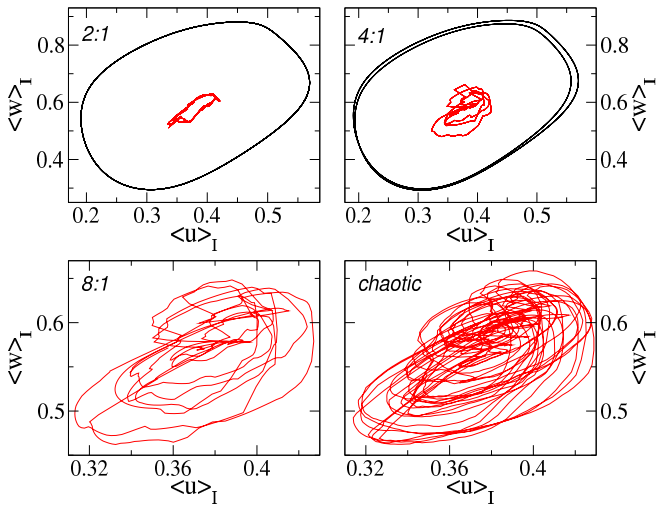}
\caption{(Color online) Dynamics of the fields $\mathbf{c}(\mathbf{r}) = (u(%
\mathbf{r}),v(\mathbf{r}),w(\mathbf{r}))$ averaged over the interfacial zone
$I(t)$, denoted as $\langle\mathbf{c}\rangle_I$, for $T_f = 1.275$ and
different values of the forcing amplitude $a>a_2^*$ corresponding to
different states of the system ($a=0.0173$ (2:1), $a=0.0155$ (4:1), $%
a=0.0147 $ (8:1), $a=0.0138$ (chaos)). The interface dynamics undergoes a
period-doubling cascade into chaos while the dynamics of the homogeneous
oscillations outside the interfacial zone (solid black curve) has only a
single period-doubling bifurcation at $a_2^2 \approx 0.0162$. The limit
cycle of the homogeneous oscillations is basically the same for the
subfigures denoted 4:1, 8:1 and chaos. }
\label{fronts_pp_2-1}
\end{figure}

There are number of interesting features of the dynamics as the system
approaches the front explosion point. Approaching $a_2^*$ from above, the
front explosion is preceded by a period-doubling bifurcation at $a^2_2$.
Fig.~\ref{fronts_pp_2-1} shows that the bifurcation occurs in the
homogeneously oscillating domains as well as in the interfacial zone. This
is in accord with the behavior of a single oscillator close to the 2:1
resonance. The phase diagram showing its states as a function of the forcing
frequency and amplitude near the 2:1 resonance is given in Figure~\ref%
{arnold_f}. As the forcing amplitude is increased at $\omega_f/\omega_0=2$
one crosses a parameter region where period doubling occurs. For the
spatially distributed system, within this period-doubled or 4:1 regime, the
interface dynamics $\langle\mathbf{c}\rangle_I = \mbox{A}_I^{-1} \int_{I(t)} %
\mbox{d}\mathbf{r} \, \mathbf{c}(\mathbf{r},t)$ undergoes further
period-doubling bifurcations into chaos before the front explodes (see Fig.~%
\ref{fronts_pp_2-1}). Here $\mbox{A}_I$ is the area of the interfacial zone.
Note that in the 4:1 regime two different types of fronts can exist, namely $%
\pi$ and $\pi/2$ fronts. Here, we have focused on the $\pi/2$ fronts since
they directly correspond to the ($\pi$) fronts present in the 2:1 regime.
Simulations show that $\pi$ fronts do not explode at $a=a_2^*$. They are
stationary even below $a_2^*$ and do not become unstable within the 4:1
regime.
\begin{figure}[htbp]
\includegraphics*[width=\columnwidth]{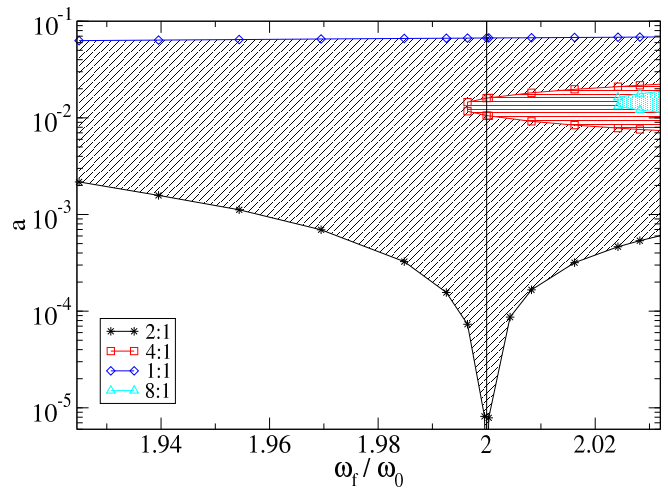}
\caption{(Color online) 2:1 Arnold tongue for a single oscillator. Note the
appearance of a period-doubling cascade within the tongue --- well below the
1:1 tongue. For $\protect\omega_f/\protect\omega_0=2$, only the first
period-doubling bifurcation is present. Its location compares very well with
the observed location for the spatially extended system. }
\label{arnold_f}
\end{figure}

Phase fronts can also be observed for random initial conditions. As long as $%
a>a_2^*$, the turbulent phase is unstable and homogeneous domains separated
by phase fronts are nucleated. If these fronts are sufficiently curved they
are not stationary but propagate. In a finite system as considered here,
this can eventually lead to the generation of a spatially homogeneous state
without any fronts. Interestingly, $\pi/2$ fronts survive much longer than $%
\pi$ fronts.

Most of the results presented in this section do not qualitatively depend on
the exact forcing frequency as long as $\omega_f$ is close to $2 \omega_0$.
Restricting ourselves to the range of forcing frequencies present in Fig.~%
\ref{arnold_f}, only the features of the dynamics as the system approaches
the front explosion point from above depend on $\omega_f$. If $\omega_f$ is
too small, the period doubling of the dynamics within homogeneous domains
and interfacial zones is absent. If $\omega_f$ is too large, a cascade of
period doubling bifurcations can be observed in the homogeneous domains as
well. In general, $a_2^*$ increases with increasing detuning from the
resonance.

\section{3:1 forcing}

\label{sec:3:1}

Near the 3:1 resonance where $T_f = T_0/3$ ($\omega_f = 3
\omega_0$) the front dynamics exhibits a number of features which
differ from those described above for the 2:1 resonance. In the
3:1 resonance regime the broken translational symmetry gives rise
to three distinct phases that differ by a phase shifts of $2 \pi /
3$; consequently, fronts separating any two of these three phases
will propagate with finite velocity $v_f$. For sufficiently large
forcing amplitudes, these propagating fronts are nearly planar
with only small transverse structure.

\begin{figure}[tbph]
\includegraphics*[width=\columnwidth]{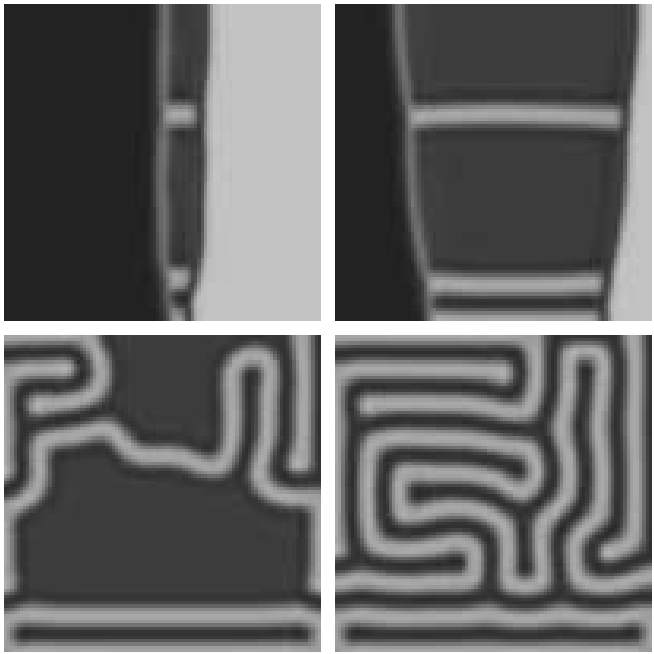}
\caption{Series of snapshots of the $u(\mathbf{r},t)$ field for $a=0.0354$
and $T_{f}=0.85$s showing the exploding front and the eventual formation of
a labyrinthine pattern. From left to right and top to bottom: $t=40\times
T_{f}$, $t=760\times T_{f}$, $t=4360\times T_{f}$, $t=10600\times T_{f}$.
Color coding as in Fig.~\protect\ref{asymp_state}. Note that due to the
initial conditions and the no-flux boundary conditions certain symmetries
persist in the pattern. }
\label{laby}
\end{figure}

As the forcing amplitude decreases, the front width $\Delta _{0}$ increases
in magnitude and, for amplitudes lower than a critical value $a_{3}^{\ast }$%
, the interfacial zone grows without bound and the front explodes. When the
interfacial zone exhibits unbounded growth, the mean velocity of the front
is zero, $v_{f}\equiv 0$. A simulation of an exploding front is shown in top
two panels of Fig.~\ref{laby}.
In contrast to the exploding front in the 2:1 resonantly forced regime, the
expanding interfacial zone does not display spatio-temporal turbulent
dynamics, instead, a labyrinthine patterns develops. This can be seen in
bottom two panels of Fig.~\ref{laby} which shows the formation of a
labyrinthine pattern after the interfacial zone fills the simulation domain.
Examination of the local dynamics indicates that the labyrinthine pattern is
locked 2:1 to the applied forcing.

\begin{figure}[tbph]
\includegraphics*[width=\columnwidth]{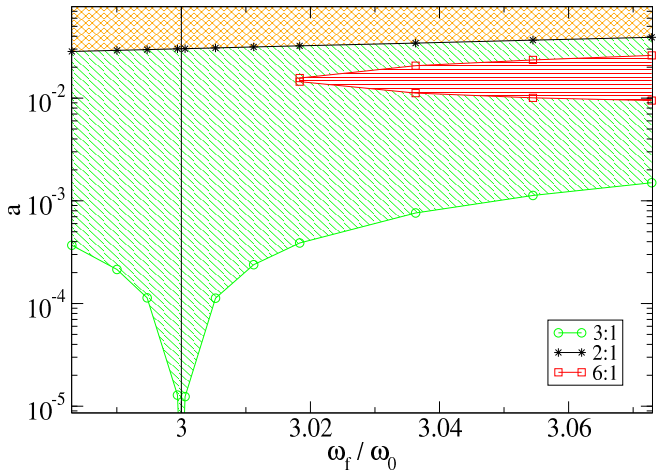}
\caption{(Color online) 3:1 Arnold tongue for a resonantly forced single
oscillator. The period-doubling bifurcation within the tongue is shown as a
region shaded with parallel lines. Note that this region is does not
intersect the 3:1 resonance line (vertical line in the figure) in contrast
to the case of 2:1 forcing. The transition from 3:1 to 2:1 locking (solid
line with *) is subcritical and the region of bistability is marked by the
orange cross pattern. }
\label{arnold3_f}
\end{figure}

Insight into origin of this behavior can be obtained by
considering the dynamics of a resonantly forced single oscillator.
The phase diagram showing its states as a function of the forcing
frequency and amplitude near the 3:1 resonance is given in
Fig.~\ref{arnold3_f}. This phase diagram has a number of features
that differ from those of the 2:1 Arnold tongue shown earlier in
Fig.~\ref{arnold_f}. The period doubling region within the Arnold
tongue does not intersect the $\omega _{f}=3\omega _{0}$ line. In
addition, there is a region, indicated by the cross pattern, where
there is bistability between 3:1 and 2:1 locking.
In the spatially extended system, the values of the forcing amplitude used
to obtain the exploding front are well within the area of bistability and
coexistence of 3:1 and 2:1 states for a single oscillator, suggesting the
possibility of appearance of a 2:1 resonantly locked state as observed in
the simulation. Of course, the single oscillator phase diagram does not
provide information on the spatial structure of the pattern and the Arnold
tongue for the spatially distributed system may display a richer structure.
Our simulation result confirms the existence of a 2:1 state with a
labyrinthine pattern within the 3:1 resonance tongue.

\begin{figure}[htbp]
\includegraphics*[width=\columnwidth]{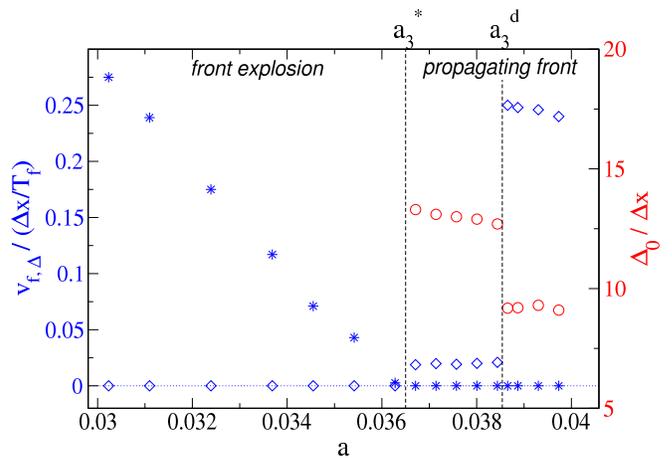}
\caption{(Color online) Plot of the mean interface width $\Delta _{0}$ (open
circles), the mean front velocity $v_{f}$ (open diamonds) and the interface
growth rate $v_{\Delta }$ (stars) as a function of the amplitude of the
forcing $a$ for $T_{f}=0.85$s. The statistical errors are less than the size
of the symbols. Three different regimes can be identified. See text for
details. }
\label{front_diagram}
\end{figure}

The front explosion can again be characterized quantitatively by plotting
the mean interfacial width, $\Delta _{0}$, the growth rate of the mean
interfacial width, $v_{\Delta }$, and the mean front velocity, $v_{f}$ as a
function of the forcing amplitude. The data in Fig.~\ref{front_diagram}
shows that there is a transition from a propagating front with $v_{f}\neq 0$
and $v_{\Delta }\equiv 0$ to an exploding front with $v_{f}=0$ and $%
v_{\Delta }>0$ at a critical value $a_{3}^{\ast }$.

\begin{figure}[htbp]
\includegraphics*[width=\columnwidth]{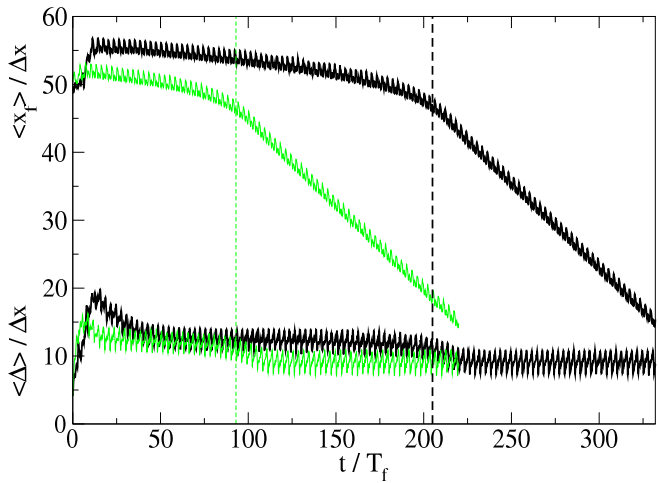}
\caption{(Color online) Temporal evolution of the average width $%
\langle\Delta\rangle$ and position $\langle x_f \rangle$ of the interfacial
zone for $T_f = 0.85$ and different values of the forcing amplitude $a>a_3^d$%
. The thin light (green) curves correspond to $a=0.03887$ and the
thick black ones to $a=0.03865$. Two regimes are visible. At short
times, a thick and slowly propagating front is present which
transforms into a thin and fast propagating front at $t^c$. The
transient time $t^c$ is shown as the dashed lines and increases
with decreasing $a$. } \label{transition}
\end{figure}

This figure also shows that there are a number of interesting
features of the dynamics in the vicinity of the front explosion.
Approaching $a_3^*$ from above, the front explosion is preceded by
a transition of the front itself. At $a_3^d$, the front velocity
suddenly drops while the width of the interfacial zone increases.
As $a$ is reduced for $a_3^* <a < a_3^d$, the width increases and
the front velocity decreases up to the transition point $a_3^*$
where $v_f=0$. The sudden change in $v_f$ and $\Delta_0$ at $a_3^d
\approx 0.038546$ is signalled by behavior of the transient time
needed to establish the interface. Figure~\ref{transition} plots
the temporal evolution of the average width and location of the
interfacial zone for $T_f = 0.85$ and different values of the
forcing amplitude. For both values of the forcing amplitude two
regimes may be distinguished in the figure. For short times a
relatively thick and slowly propagating front is observed which
persists for a transient time $t^c$ indicated by the dashed lines
in the figure. For times $t> t^c$, the front transforms into a
faster moving thin front. The evolution of the interfacial zone
for $t<t^c$ is very similar to that observed for $a<a_3^d$. As $a$
approaches $a_3^d$ from above, the transient time $t^c$ increases
as shown in the figure. For $a<a_3^d$, no evidence for a finite
$t^c$ has been found. Thus, these data suggest that transient time
$t^c$ diverges at $a_3^d$ leading to a distinctly different front
types below and above $a_3^d$.

Another interesting feature of the dynamics in the vicinity of the
front explosion is that, approaching $a_3^*$ from above, the front
explosion is preceded by a sequence of period-doubling and inverse
period-doubling bifurcations in the dynamics of the interfacial
zone characterized by the behavior of
$\langle\mathbf{c}\rangle_I(t)$. Depending on the value of $a$,
$\langle\mathbf{c}\rangle_I(t)$ is locked either 3:1 or 6:1 to the
forcing; however, there is no period-doubling bifurcation of the
homogeneous oscillations outside the interfacial zone in contrast
to the case of 2:1 forcing. This observation is expected, based on
the phase diagram for a single oscillator given in
Fig.~\ref{arnold3_f} which does not show a period-doubling
bifurcation close to the 3:1 resonance.

Finally, we observe that if random initial conditions are chosen,
labyrinthine patterns form for $a<a_{3}^{\ast }$ while homogeneous
domains locked 3:1 to the forcing and separated by phase fronts occur for $%
a>a_{3}^{\ast }$. In the latter case, due to the propagation of the fronts,
these inhomogeneous states generally evolve to homogeneous asymptotic states
in a finite system.

\section{Conclusions and Discussion}

The results presented in this paper show that while front \textquotedblleft
explosions" occur in a realistic model for catalytic CO surface oxidation
reactions, a number of new phenomena are observed. In the 3:1 regime of the
resonantly forced CGLE,  when the underlying unforced system is
Benjamin-Feir unstable \cite{faradaypaper,physicaDpaper}, and also for the
period-3 piecewise linear coupled maps \cite{klop,klp}, propagating fronts
have a turbulent interfacial zone separating homogeneous resonantly-locked
states. The interfacial zone grows in this case as a power law when the
magnitude of the forcing amplitude is decreased. In the CO oxidation
reaction, on the other hand, the interfacial zone in the propagating front
is much more structured and front explosion leads to the formation of a
labyrinthine pattern with 2:1 locking. Investigations of the front explosion
dynamics in the 2:1 regime of the CGLE with the Benjamin-Feir instability
reveal propagating Bloch fronts with a turbulent interfacial zone which
grows without bound beyond a certain critical value of the forcing amplitude
\cite{chris}. In both the 2:1 CGLE and in period-2 coupled maps,
poorly-characterized front explosions of stationary fronts have also been
observed. In the 2:1 resonantly forced CO surface oxidation reaction, the
stationary front separating the two homogeneous resonantly locked states has
a turbulent interfacial zone. As the forcing amplitude is decreased, beyond
a critical value the interfacial zone grows without bound leading to a
defect-mediated turbulent state.

Although spatially-resolved experiments with periodic forcing of
the CO oxidation reaction have already been performed
\cite{BertramJPC}, the behavior of fronts was not studied. To
create a front, special initial conditions are needed, which were
not implemented in these experiments where forcing was always
applied starting from the turbulent initial state. An additional
difficulty was that, in the experiments, the frequency of uniform
oscillations in absence of forcing could not be exactly measured,
because such oscillations were unstable and led to turbulence. In
these experimental investigations, the attention was focused on
qualitative properties of different patterns induced by forcing.
Remarkably, irregular stripe patterns with the 2:1 locking were
found then for high forcing frequencies corresponding to 3:1
forcing \cite{BertramJPC}. Such irregular stripe patterns were
interpreted as labyrinthine patterns affected by strong surface
anisotropy. In the present theoretical study, we have indeed found
that the 2:1 locked labyrinthine patterns should exist under 3:1
resonant forcing in the CO oxidation system.

Our analysis of front explosions provides a theoretical framework
for future experimental studies of the CO oxidation reaction under
resonant periodic forcing. Generally, it provides an example of
the front explosion phenomena for a system with non-harmonic
oscillations and more complex dynamics, different from the ideal
situation described by the complex Ginzburg-Landau equation.

Acknowledgements: The research of RK was supported in part by a
Humboldt Research Award of the Alexander von Humboldt-Stiftung
(Germany), and the Natural Sciences and Engineering Research
Council of Canada.

\end{document}